\documentclass[showpacs]{revtex4}
\usepackage{graphicx,amsmath,rotating,longtable}

\begin{document}
\title{Active motions of Brownian particles in a generalized energy-depot model}
\author{Yong Zhang} \email{xyzhang@phya.yonsei.ac.kr}
\author{Chul Koo Kim} \email{ckkim@yonsei.ac.kr}
\address{Institute of Physics and Applied Physics, Yonsei
University, Seoul 120-749, Korea}
\author{Kong-Ju-Bock Lee} \email{kjblee@ewha.ac.kr}

\address{Department of Physics, Ewha
Womans University, Seoul 120-750, Korea}
\address{School of
Physics, Korea Institute for Advanced Study, Seoul 130-722, Korea}
\date\today

\begin{abstract}
We present a generalized energy-depot model in which the
conversion rate of the internal energy into motion can be
dependent on the position and the velocity of a particle. When the
conversion rate is a general function of the velocity, the active particle
exhibits diverse patterns of motion including a braking mechanism and a stepping motion.
The phase trajectories of the motion are investigated in a systematic way.
With a particular form of the conversion rate dependent on the position and velocity,
the particle shows a spontaneous oscillation characterizing a
negative stiffness.  These types of active behaviors are compared with the similar phenomena observed
in biology such as the stepping motion of molecular motors and the amplification in hearing mechanism.
Hence, our model can provide a generic understanding of the
active motion related to the energy conversion and also a new control mechanism
for nano-robots. We also
investigate the noise effect, especially on the stepping motion
and observe the random walk-like behavior as expected.
\end{abstract}
\maketitle

\section{Introduction}

Active motion is a phenomenon found ubiquitously in nature ranging
from the colony organization of microorganism, biological springs,
molecular motors, nano-robotics~\cite{SCIENCE_v288_95,
SCIENCE_v308_1308,BMB_v69_539,PRE_v76_011919,NM_v6_235},  a flocking
of fish and birds to swarming of small
insects~\cite{BB_v202_296,AnnPhys_v318_317}. All of these types of
active motions need processes of energy supply,  conversion
to motion, and resupply for repeated motions. To describe
these processes, an energy depot model was introduced by
Schweitzer {\it et. al}~\cite{PRL_v80_5044,Schweitzer}. In the energy
depot model, supplied energy is able to induce effectively a
negative friction in a certain range of velocity yielding an
active motion. This model was successfully applied to a wide
variety of active motions~\cite{BMB_v69_539,PRE_v76_011919,
PRL_v80_5044,Schweitzer,PhysicaA_v273_294,EPJB_v14_157,ActaPP_v39_1251}
mainly assuming that
the energy conversion rate into the motion depends only on the
quadratic form of the velocity. This success of the quadratic energy
conversion rate model immediately raises the possibility that other diverse active
motions can be induced, when the energy conversion rate contains various
terms in velocity including the quadratic one.

Another motivation for the present study originates from the
fields of nano-robotics and artificial
molecular motions~\cite{NM_v6_235,SCIENCE_v318_1155}.
Rapid developments in these area require diverse controlling
technology for energy conversion for intended purpose.
Although it is still at a primitive stage at present,
control of the energy conversion will become fine-tuned in the near future
so that any desired motion could be selectively achieved.
Motivated by these considerations, we consider various
forms of the conversion rate of the internal energy into a motion
in this paper. Hence our work provides a correlation between the
form of energy conversion and the active motion.

In the following sections, we introduce our generalized energy
depot model in detail and consider various forms of the conversion
rate which could induce diverse active motions including a braking mechanism applicable to
nonlinear compressive behavior of basilar membrane in the inner
ear~\cite{Zhang}, a directed stepping motion shown in the
molecular motor
systems~\cite{NATURE_v365_721,Cell_v126_335,QRBP_v40_87}, or a
phenomenon of negative stiffness for oscillatory system such as
the hair bundle in mammals~\cite{PNAS_v102_16996}. Concluding
remarks will follow.

\section{Generalized Energy Depot Model}

A Brownian particle moving in an external force $f(x)$ is governed
by the Langevin equation,
\begin{eqnarray}
m\frac{dv}{dt} = -\mu_0 v + f(x) + \sqrt{2k_BT \mu_0}\zeta (t),
\label{la}
\end{eqnarray}
where $m$ is the mass, $v$ the velocity of the particle, $\mu_0$
the friction coefficient, $T$ the temperature, and the white noise,
$\zeta(t)$ satisfies
$\langle \zeta(t)\rangle = 0$ and $\langle \zeta_i(t)\zeta_j(t^{\prime})\rangle
= \delta_{ij}\delta(t-t^{\prime})$. The passive motion can be converted to the active motion
when the internal energy depot is introduced. The internal energy
of the depot, $e(t)$, is regarded as an additional degree of
freedom for each particle. The energy balance equation for the
depot can be described as~\cite{Schweitzer}
\begin{eqnarray}
\alpha \frac{de(t)}{dt} = q(x) - ce(t) - d(x,v)e(t),
\label{depot}
\end{eqnarray}
where $\alpha$ is the timescale of relaxation of the depot, $q$
the rate of the energy influx to the depot, $c$ the rate of energy
dissipation of the depot, $d(x,v)$ the conversion rate of the
internal energy into motion. $\alpha=1$ means that the depot reacts with a
time lag and $\alpha\rightarrow 0$ that the depot
adapts very fast (adiabatic approximation).
We note that the conversion rate of the internal energy into motion can be,
in general, a function of both velocity and position of the particle,
which contains various combinations of the two
variables~\cite{PRL_v80_5044,Schweitzer,
NJP_v9_136,PLB_v659_447}. In order to understand the roles of
individual terms, we expand $d(x,v)$ as follows,
\begin{eqnarray}
d(x,v) = \sum_{i,j=0}^\infty a_{i,j} x^i v^j. \label{ansaz}
\end{eqnarray}
We can choose $a_{0,0} = 0$ since the effect of constant
conversion rate can be merged into the constant dissipation
rate.

Consequently, the Langevin equation for an active Brownian
particle is written as
\begin{eqnarray}\label{actLE}
\begin{aligned}
m \frac{dv}{dt} &= -\mu_0 v + f(x) + F_{active} +
\sqrt{2k_BT \mu_0}\zeta(t), \\
 &= -\mu(x,v)v
+ f(x) + \sqrt{2k_BT\mu}\zeta (t),
\end{aligned}
\end{eqnarray}
where $F_{active} = d(x,v)e(t)/v$ and $\mu(x,v) = \mu_0 -
d(x,v)e(t)/v^2$. The above equations are intuitive in a sense that
the Brownian particle is governed by an extra active force or that
the friction is modified to depend on the space and velocity, when
external energy is supplied. It is noticeable that the effective
friction can be negative or larger than the normal friction
$\mu_0$. The effective friction will be discussed in the next
section in detail.

In this paper, $q$ and $c$ are assumed to be constant for
simplicity and main focus is given on the effect of the conversion
rate. For this purpose, we consider the conversion rate
$d(v)$ depending only on the velocity  first and  $d(x,v)$ in a
special form later. Even when only velocity-dependent conversion rate
is considered, diverse properties can be discussed such as the
motion with a braking mechanism, the stability of fixed points, and
a stepping motion. A special form of $d(x,v)$ is chosen to
describe a possibility of a negative stiffness.

\section{Motion with A Braking Mechanism}

The active Brownian particle with $d(v)\sim v^2$ has been
previously treated in detail and it is shown that the friction can
be effectively negative when the particle moves slowly and
increases only up to $\mu_0$ as the particle moves
faster~\cite{PRL_v80_5044,Schweitzer}.
Hence, for an active particle with $d(v)\sim
v^2$, speed can be increased without any limit. Since such an excessive
speed can damage living organisms, it may be possible
that a living organism is equipped with a protective mechanism to
prevent damage from excessive movement or energy pumping,  especially
in underdamped oscillating systems. Also, in nano-robotics, adoption of
this type of protective
mechanism may be not only helpful to control the movement, but also
essential to safeguard the mechanism from overdriving. Thus, study on
contributions from higher order velocity terms are highly desirable.

Motivated from these arguments, we first study a symmetric form
including up to a fourth order term in the conversion rate of the
internal energy into motion. Hence,
\begin{eqnarray}
d(v) = a_{0,2} v^2 + a_{0,4}v^4, \label{add}
\end{eqnarray}
where the positive (negative) $a_{0,4}$ increases (decreases) the
kinetic energy of the particle. Since we are interested in the
braking mechanism,  only the negative case will be treated. By
introducing a critical velocity, $v_c = \sqrt{a_{0,2}/|a_{0,4}|}$,
the conversion rate is rewritten as
\begin{eqnarray} d(v) = a_{0,2} v^2 \left(1 -
\frac{v^2}{v_c^2}\right). \label{f1}
\end{eqnarray}
It implies that the energy depot reabsorbs the kinetic energy when
the velocity goes beyond the critical velocity. This mechanism is
analogous to the regenerative-brake system in automobiles
and electric vehicles~\cite{ACC_v4_3129}. It is shown that
this regenerative-braking can not only control the motion more effectively,
but also save energy for operation.
Thus, we name
this fourth-order model as a Brownian energy depot model with a
braking mechanism.
\begin{figure}
\begin{center}
\includegraphics*[width=1.0\columnwidth]{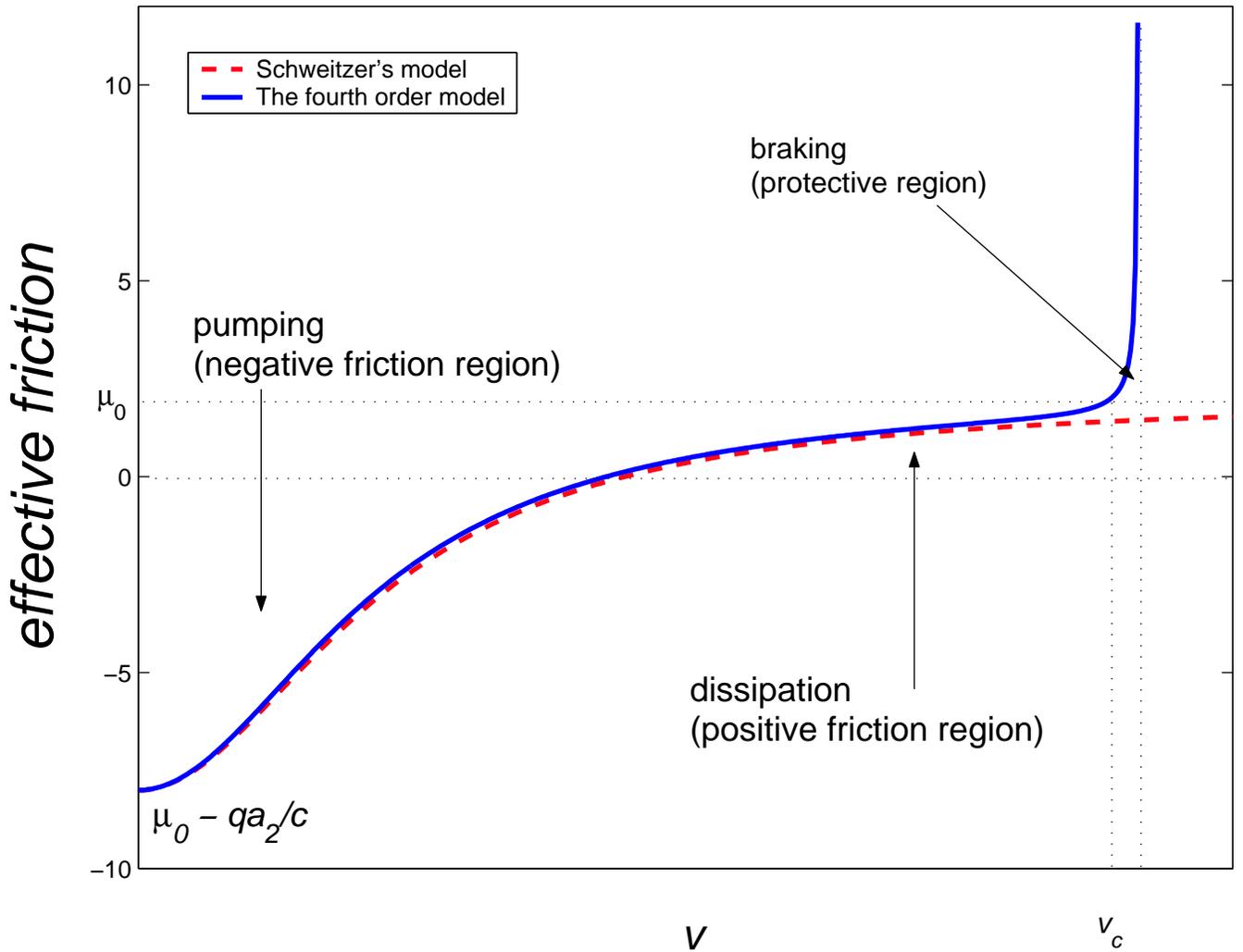}
\end{center}
\caption{(color online) An illustration of the effective friction
as a function of the velocity. Solid line shows the effect of the
fourth order term in the conversion rate, compared with the
Schweitzer's result in dashed line. $q = 10$, $\mu_0 = 2.0$, $v_c
= 5.0$ and $a_{0,2}/c = 1$ are used for the numerical
calculation.} \label{friction}
\end{figure}

To describe the effect of the fourth order term in the conversion
rate, let us consider an adiabatic approximation,
$\alpha\rightarrow 0$, in which the energy depot adapts very fast.
The adiabatic solution of the internal energy of depot
yields~\cite{Schweitzer}
\begin{eqnarray}
e(t) = \frac{q}{c + a_{0,2}\left(1-v^2/v_c^2\right) v^2}.
\label{e1}
\end{eqnarray}
Hence the effective friction coefficient becomes
\begin{eqnarray}\label{mu}
\begin{aligned}
\mu(v) &= \mu_0 - \frac{d(v)}{v^2}e(t) \\
 &= \mu_0 -
\frac{qa_{0,2}(1-v^2/v_c^2)}{c + a_{0,2}(1 - v^2/v_c^2)v^2}.
\end{aligned}
\end{eqnarray}
Note that $\mu(v_c)=\mu_0$. The effective friction describes three
different situations ; (i) pumping where $\mu(v) < 0$, (ii)
dissipation where $0 < \mu(v) < \mu_0$, and (iii) protection where
$\mu_0 < \mu(v)$. The particle cannot speed up when the velocity
exceeds the critical velocity in this model, since the fourth order
term in the conversion rate induces sufficiently large friction to
the particle. Superficially, the present model appears to be similar
to the Rayleigh model~\cite{Rayleigh}, but with different
interpretations.

We believe that the newly introduced braking mechanism may offer a
plausible way to discuss any underdamped motions and protection from
any overreacting motion in living organisms and nano-robotics.
Recently by adopting this mechanism to the active oscillation of
the basilar membrane (BM) in the mammalian ear, we could
successfully explain the known experimental results and the noise
amplification in cochlea~\cite{Zhang}.

\section{Stability of Fixed Points}

So far, we have restricted the conversion rate $d(v)$ to be symmetric in
the velocity. However, the active particles generally possess
polarities in their motions~\cite{NM_v6_235,SCIENCE_v318_1155,
NATURE_v365_721,Cell_v126_335,QRBP_v40_87}. The polarity of the
motion is originated from the external force or/and the asymmetric
conversion rate. To discuss the generic polarity of the motion, in
this section, we consider a general form of $d(v)$,
\begin{eqnarray}\label{dv1}
d(v) &= a_{0,1}v + a_{0,2}v^2 + a_{0,3}v^3 + a_{0,4}v^4 + \cdots.
\end{eqnarray}
It is expected that inclusion of these odd terms allows the active
particles perform polarized motions as observed in molecular motors such as
kinesin and dynein. Now, it is worth to analyze the deterministic dynamics
in the
$\{v,e\}$ phase space which is governed by
\begin{eqnarray}\label{ds}
\begin{aligned}
\dot{v} &= -\frac{\mu_0}{m}v + \frac{f}{m} + \frac{d(v)}{mv}e , \\
\dot{e} &= q - ce - d(v)e,
\end{aligned}
\end{eqnarray}
where the depot takes time to be filled with energy ($\alpha=1$).
It is well known that long time behavior of a two-dimensional
continuous dynamical system possesses only fixed points and
limit cycles~\cite{Strogatz}. Although it is not possible to
describe the solution analytically, local stability for existing
fixed points can be analyzed through the Jacobian matrix and its
eigenvalues~\cite{Schweitzer,Strogatz}. In the following, we discuss the
stability of the fixed points obtained numerically for various
values of $a_{i,j}$. For the numerical calculation, $m=1$, $q =
10$, $c = 0.01$ and $\mu_0 = 20$ are used.

\begin{table}
\caption{Bifurcation with a variation of $a_{0,3}$, when
$d(v)=a_{0,1}v+a_{0,2}v^2+a_{0,3}v^3$, $a_{0,1} = 0.0002$,
$a_{0,2} = 2.0$, and $f=0$. $(+/-)$ denotes positive/negative
velocity.}
\label{T1}
\begin{tabular}{@{} l l}
\hline\hline
 Range of $a_{0,3}$ & Classification of the fixed point(s) \\
\hline
 $a_{0,3} \leq -0.93$ & $\cdot$ one unstable focal fixed point (-) \\
        & $\cdot$ one saddle point (-) \\
        & $\cdot$ one stable fixed point (+) \\ \\

 $-0.93 < a_{0,3} < -0.16$ & $\cdot$ one unstable focal fixed point (-) \\
        & $\cdot$ one saddle point (-) \\
        & $\cdot$ one stable focal fixed point (+)\\ \\

 $-0.16 \leq a_{0,3} < 0.16$ & $\cdot$ one stable focal fixed point (-) \\
        & $\cdot$ one saddle point (-) \\
        & $\cdot$ one stable focal fixed point (+) \\ \\

 $0.16 \leq a_{0,3} < 0.93$ & $\cdot$ one stable focal fixed point (-) \\
        & $\cdot$ one saddle point (-) \\
        & $\cdot$ one unstable focal fixed point \\
          & \, with a limit cycle around it (+) \\ \\

 $a_{0,3} \geq 0.93$ & $\cdot$ one stable fixed point (-) \\
        & $\cdot$ one saddle point (-) \\
        & $\cdot$ one unstable focal fixed point \\
          & \, with a limit cycle around it (+) \\
\hline
\end{tabular}
\end{table}

Table~\ref{T1} shows the bifurcation with the variation of
$a_{0,3}$ when we consider the conversion rate only up to the
third order of velocity and there is no external force. There is
always at least one stable fixed point or stable focal fixed point in the $\{v,e\}$
space. The trajectories initiated around the stable fixed point or stable
focal fixed point are finally localized at this fixed point. When
$-0.16 < a_{0,3} < 0.16$, there are two stable focal points. On
the other hand, a limit cycle appears  if $a_{0,3}\ge 0.16$.
\begin{figure}
\begin{center}
\includegraphics*[width=1.0\columnwidth]{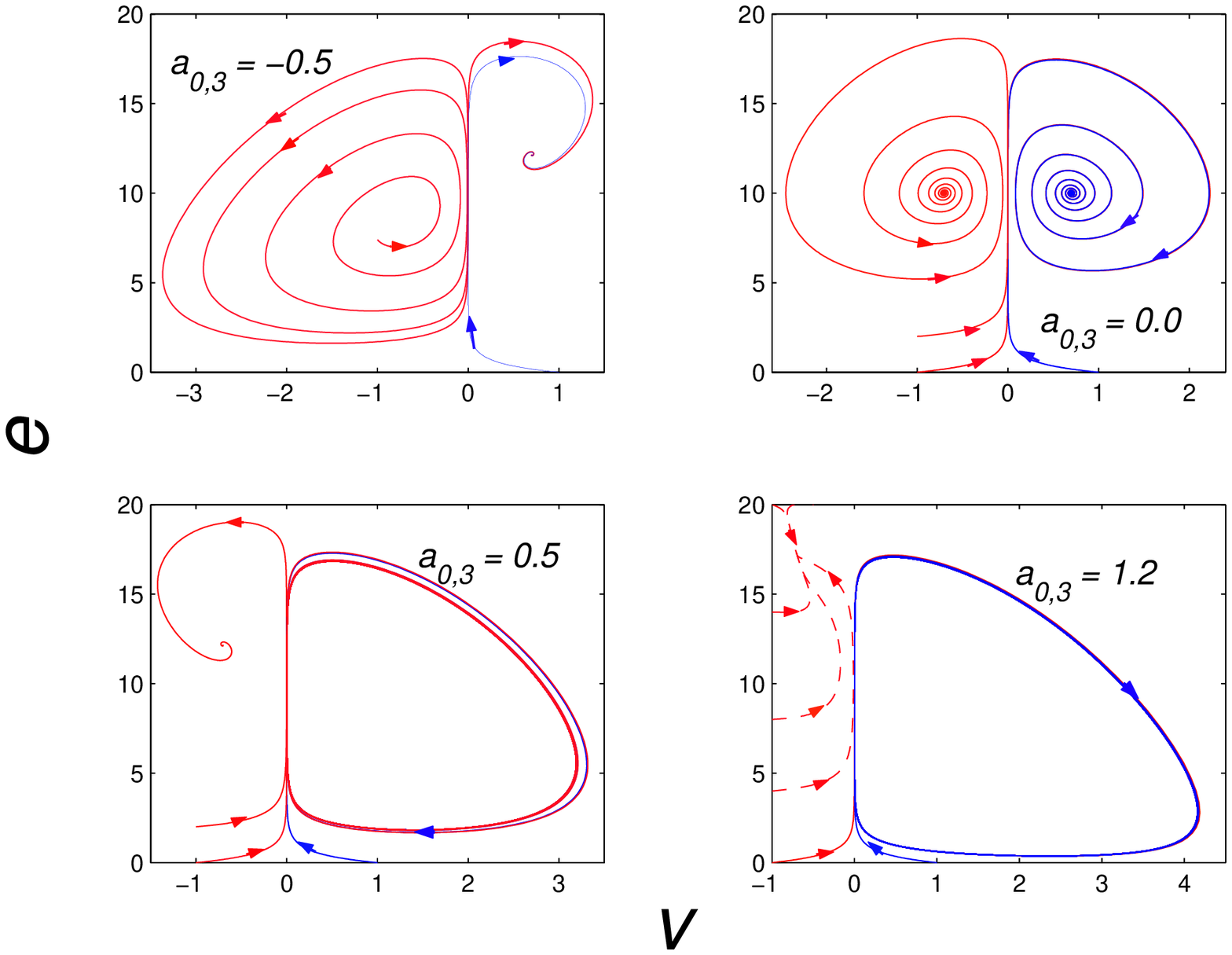}
\end{center}
\caption{Trajectories in the phase space, $\{v,e\}$ for several
values of $a_{0,3}$, when $a_{0,1} =0.0002$, $a_{0,2} = 2.0$, and
$f = 0$. Note that a limit cycle appears only if $a_{0,3}\ge
0.16$.} \label{a3}
\end{figure}
In Fig.~\ref{a3}, we show several phase trajectories on the phase
space, \{$v$,$e$\} for several values of $a_{0,3}$ which describe
different classifications of fixed points as listed in
Table~\ref{T1}. Here, we note that overall
behavior of this system is not much sensitive to variations of  $a_{0,1}$.
\begin{figure}
\begin{center}
\includegraphics*[width=0.7\columnwidth]{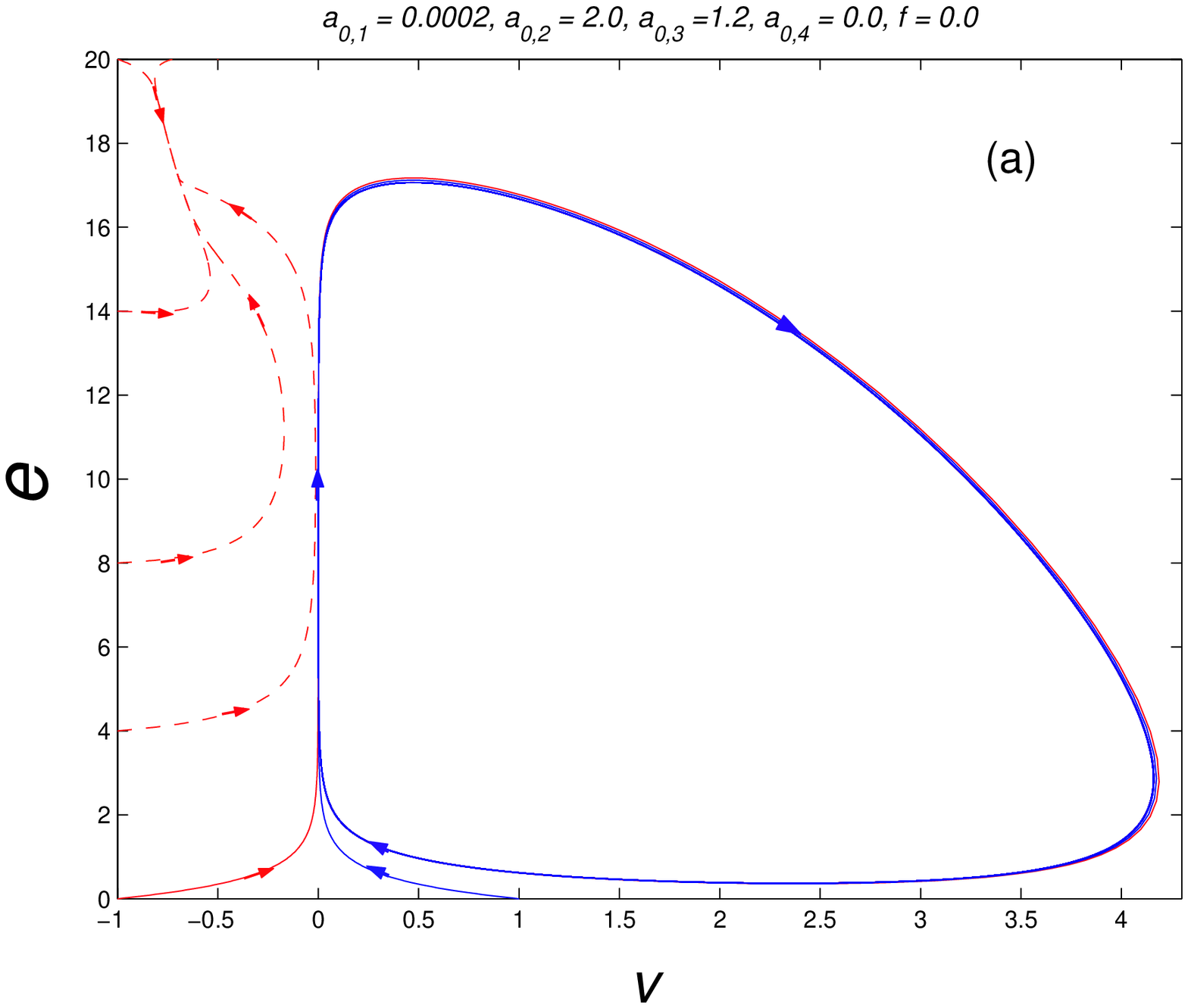}
\includegraphics*[width=0.7\columnwidth]{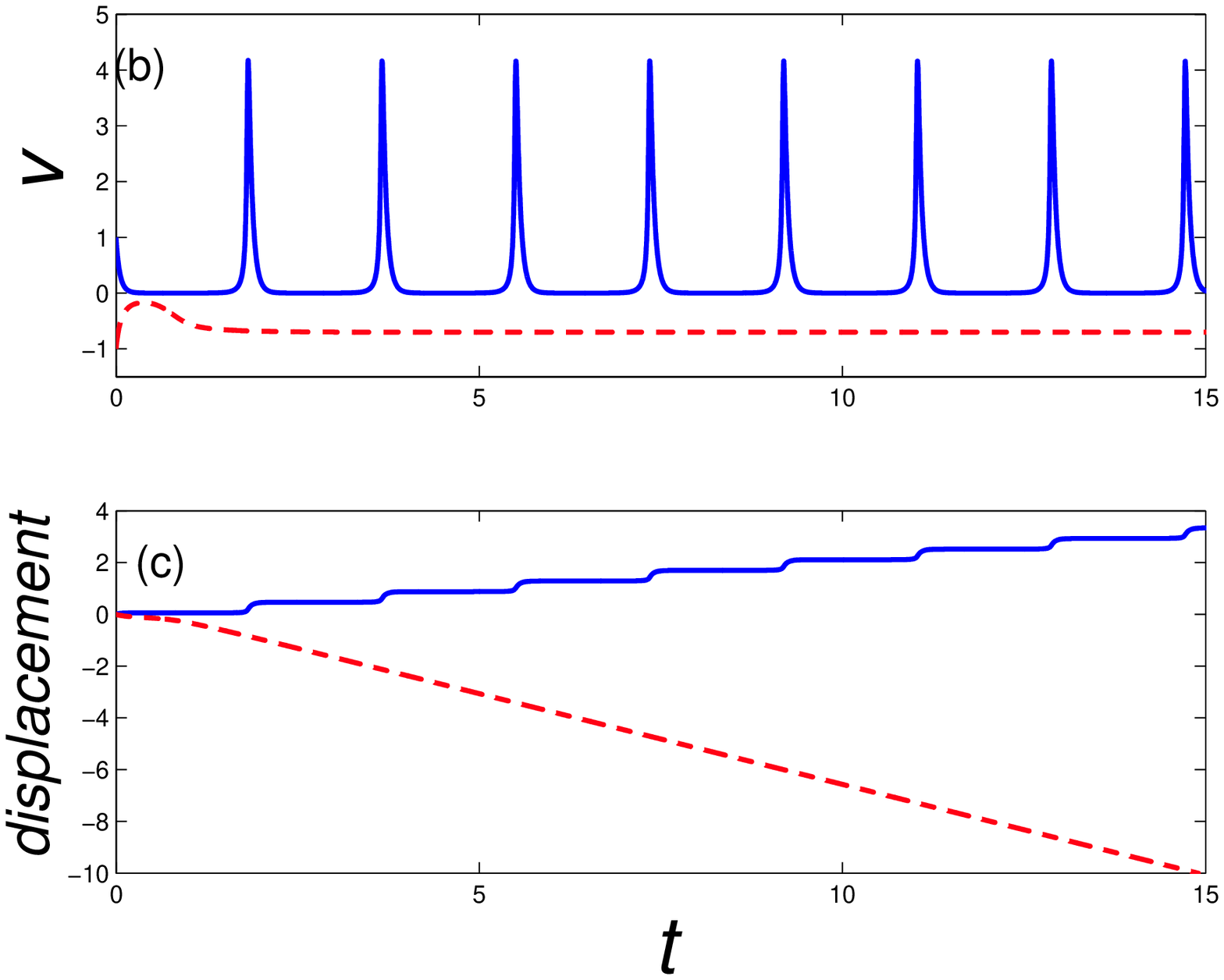}
\end{center}
\caption{(color online) (a) Trajectories in the phase space,
$\{v,e\}$ when $a_{0,1} =0.0002$, $a_{0,2} = 2.00$, $a_{0,3}=1.2$,
and $f = 0$. (b) When the initial velocity is in the positive
(negative) direction, the velocity has a steady jerky (smooth) pattern
in time. (c) Displacement of the particle is step-like (smooth)
for the limit cycle (stable fixed point).} \label{jerky}
\end{figure}

To see the pattern of motion for the limit cycle, we replot the
phase diagram for $a_{0,3}=1.2$ in Fig.~\ref{jerky}(a) with
corresponding $v(t)$ and $x(t)$ in Fig.~\ref{jerky}(b) and (c),
respectively. It shows that the quasi-triangle shaped limit cycle corresponds to a
step-by-step movement of the particle (solid line) while the
stable fixed point does to a smooth movement (dashed line). The
direction of motion depends on the initial velocity. We will
discuss this stepping motion in detail later because it may be applicable
to the molecular motor systems in biology.

Now, we include the fourth order term in the conversion rate,
$d(v)$ and analyze the classification of the fixed points as
$a_{0,4}$ is varied.
\begin{eqnarray}
d(v) = a_{0,1}v + a_{0,2}v^2 + a_{0,3}v^3 +a_{0,4}v^4. \label{ss2}
\end{eqnarray}
For the numerical analysis, we use  $a_{0,1} = 0.0002$,
$a_{0,2} = 2.0$, $a_{0,3}=1.2$, and $f=0$. We will discuss the
effect of the external force later.

\begin{table}
\caption{Bifurcation with a variation of $a_{0,4}$ when
$d(v)=a_{0,1}v+a_{0,2}v^2+a_{0,3}v^3+a_{0,4}v^4$, $a_{0,1} =
0.0002$, $a_{0,2} = 2.0$, $a_{0,3}=1.2$, and $f=0$. $(+/-)$
denotes positive/negative velocity.}
\label{T3}
\begin{tabular}{@{} l l}
\hline\hline
 Range of $a_{0,4}$ & Classification of the fixed point(s) \\
\hline
$a_{0.4} < -1.93$ & $\cdot$ one stable fixed point (-) \\
        & $\cdot$ one saddle point (-) \\
        & $\cdot$ one stable fixed point (+) \\ \\

$-1.93 \leq a_{0,4} < -0.69$ & $\cdot$ one stable fixed point (-) \\
        & $\cdot$ one saddle point (-) \\
        & $\cdot$ one stable focal fixed point (+) \\ \\

$-0.69 \leq a_{0,4} < 0.25$ & $\cdot$ one stable fixed point (-) \\
        & $\cdot$ one saddle point (-) \\
        & $\cdot$ one unstable focal fixed point \\
         & \, with a limit cycle around it (+) \\ \\

$0.25 \leq a_{0,4} < 0.92$ & $\cdot$ one stable focal fixed point (-) \\
        & $\cdot$ one saddle point (-) \\
        & $\cdot$ one unstable focal fixed point \\
         & \, with a limit cycle around it (+) \\ \\

$0.92 \leq a_{0,4} < 2.89$ & $\cdot$ one unstable focal fixed point (-) \\
        & $\cdot$ one saddle point (-) \\
        & $\cdot$ one unstable focal fixed point \\
         & \, with a limit cycle around it (+) \\ \\

$a_{0,4} \geq 2.89$ & $\cdot$ one unstable fixed point (-) \\
        & $\cdot$ one saddle point (-) \\
        & $\cdot$ one unstable focal fixed point \\
         & \, with a limit cycle around it (+) \\
\hline
\end{tabular}
\end{table}

\begin{figure}
\begin{center}
\includegraphics*[width=0.7\columnwidth]{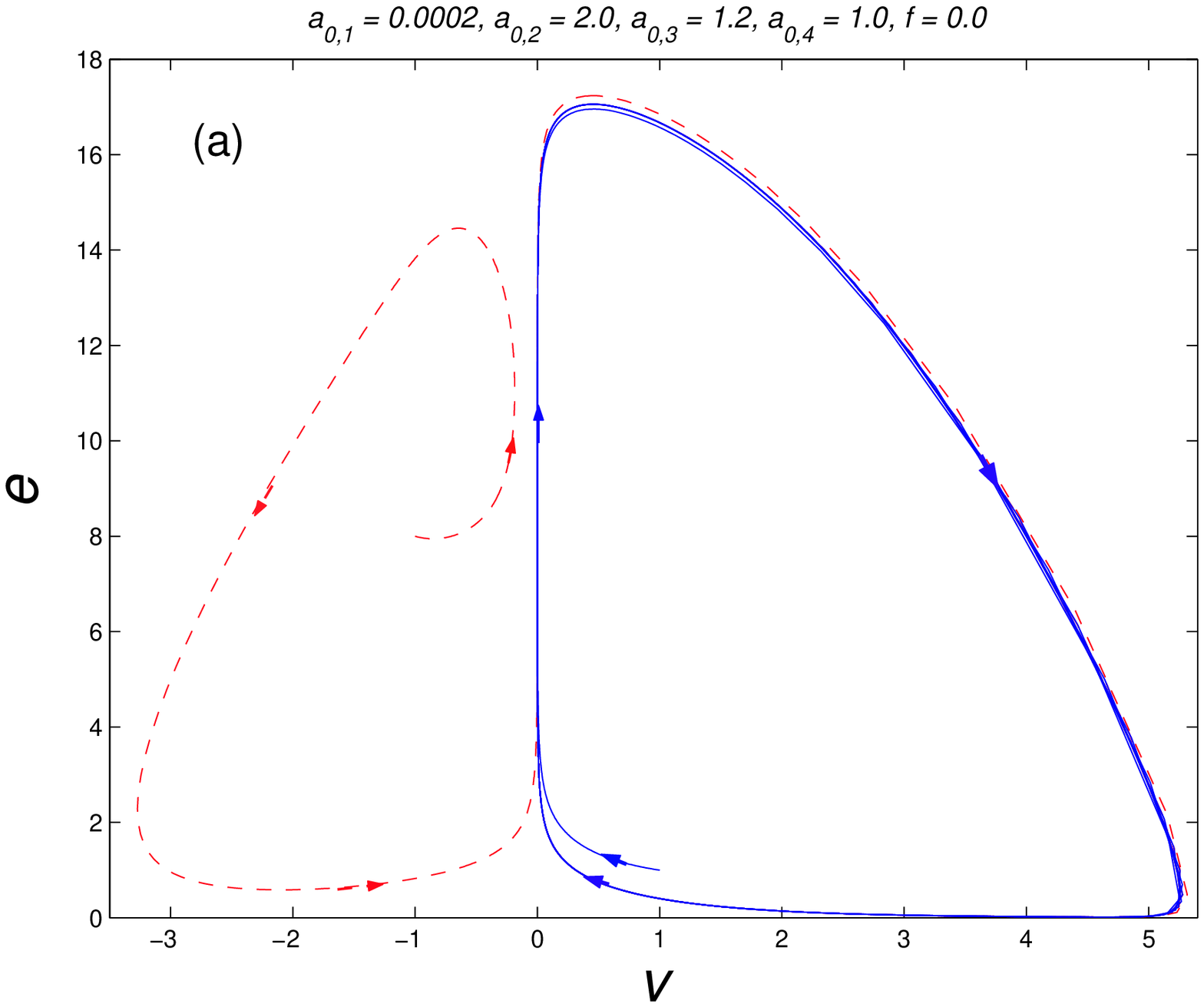}
\includegraphics*[width=0.7\columnwidth]{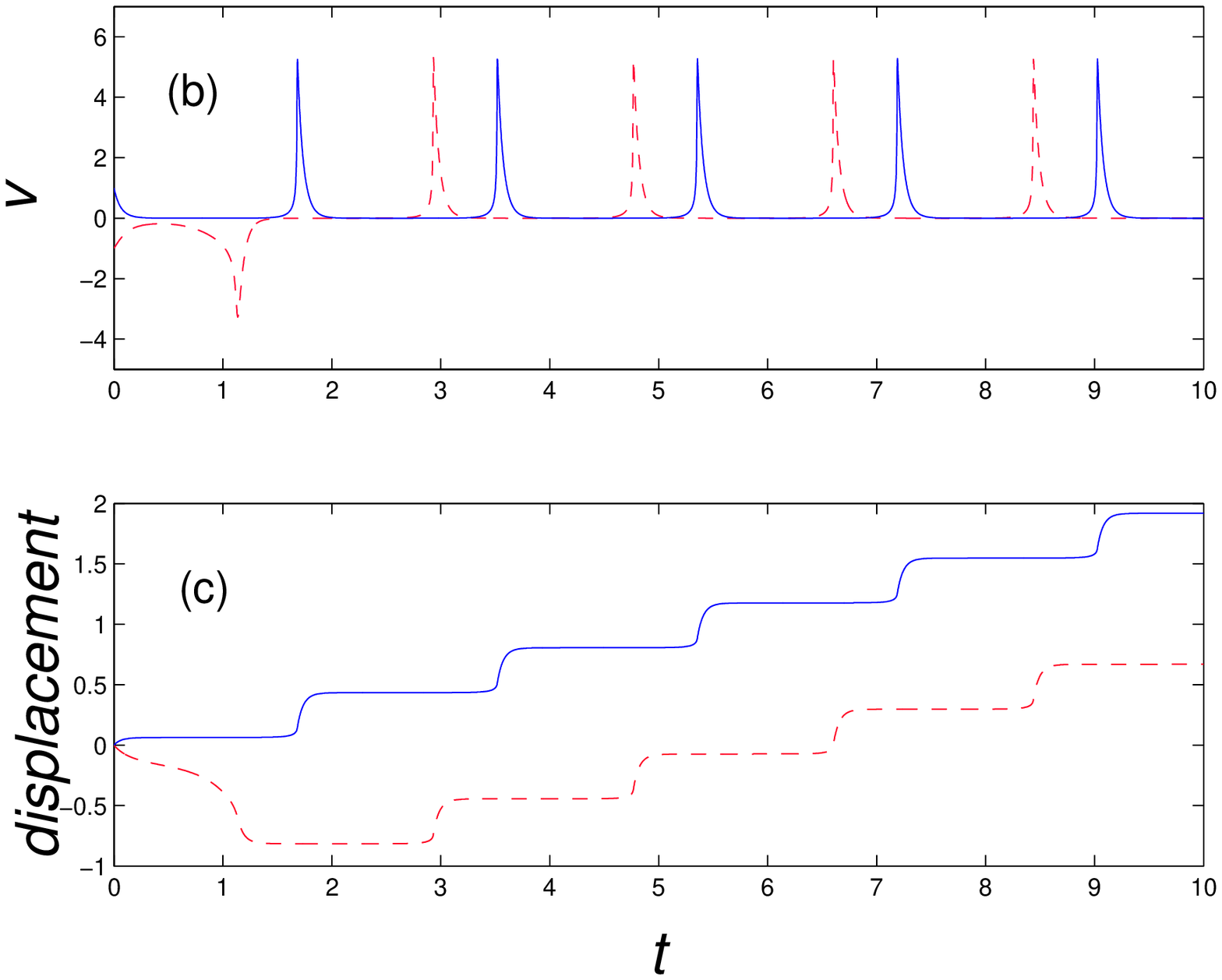}
\end{center}
\caption{(color online) Trajectories when $a_{0,1} = 0.0002$,
$a_{0,2} = 2.0$, $a_{0,3} = 1.2$, $a_{0,4} = 1.0$, and $f=0$. (a)
The phase diagram of the particle in the $\{v,e\}$ space. As shown in
Table~\ref{T3}, There is one limit cycle around the fixed point
with the positive velocity. (b) The velocity is jerky regardless
of the direction of the initial velocity. (c) The particle shows
stepping motions to the positive direction even for the initial
motion of negative velocity.} \label{a1-4}
\end{figure}
In Table~\ref{T3}, the bifurcations and
corresponding nature of fixed points for different values of $a_{0,4}$
are listed. When $a_{0,4} <
-0.69$, the particle moves smoothly regardless of the initial
velocity. As $a_{0,4}$ increases, the motion starts to depend on
the initial state of the particle. When $-0.69 \leq a_{0,4} <
0.92$, the particle shows  a directed stepping motion or a
smooth motion depending on the sign of the initial
velocity. Interestingly, when $a_{0,4} > 0.92$, the
particle moves stepwise in the positive direction eventually
regardless of its initial velocity. We plot this behavior in Fig~\ref{a1-4}, when
$a_{0,4}=1.0$.

So far, we have not included any external forces to the system.
However, it has been shown that the active particle is able to
show a directed motion, without any external forces, depending on
the structure of the conversion rate of the internal energy into
the motion. Another interesting feature is that the directed
motion could be stepwise. Such a directed stepwise motion appears in
the motion of molecular motors in biology. The molecular motors
usually carry external loads dictated by their functions. In order to
simulate this load-carrying capability of the molecular motors, we now apply
an external force, $f$, to the active particle. To analyze the
bifurcation as $f$ varies, we assume that $f$ remains constant during the motion
and use that $a_{0,1} = 0.0002$, $a_{0,2} = 2.0$, $a_{0,3} = 1.20$,
and $a_{0,4} = 1.0$. Table~\ref{T4} shows the classification of the
fixed points for various ranges of $f$, where a positive force
pulls the particle to the positive direction and a negative force
acts as a load against the positive direction of motion.

\begin{table}
\caption{Bifurcation with a the variation of the external force when
$a_{0,1} = 0.0002$, $a_{0,2} = 2.0$, $a_{0,3} = 1.2$ and $a_{0,4} =
1.0$. $(+/-)$ denotes positive/negative velocity.} \label{T4}
\begin{tabular}{@{} l l}
\hline\hline
 Range of $f$ & Classification of the fixed point(s) \\
\hline
 $f \leq -2.77$ & $\cdot$ one stable focal fixed point (-) \\
         & $\cdot$ one saddle point (+) \\
         & $\cdot$ one unstable fixed point (+)\\ \\

 $-2.77 < f < -0.89$ & $\cdot$ one stable focal fixed point (-) \\
        & $\cdot$ one saddle point (+) \\
         & $\cdot$ one unstable focal fixed point (+) \\ \\

 $-0.89 \leq f < -0.85$ & $\cdot$ one stable focal fixed point \\
    &\, with a limit cycle around it (-) \\
        & $\cdot$ one saddle point (+)\\
         & $\cdot$ one unstable focal fixed point (+) \\ \\

 $-0.85 \leq f < -0.2$ & $\cdot$ one unstable focal fixed point \\
    &\, with a limit cycle around it (-) \\
        & $\cdot$ one saddle point (+) \\
         & $\cdot$ one unstable focal fixed point (+) \\ \\

 $ f = -0.2$ & $\cdot$ one unstable focal fixed point \\
    &\, with a limit cycle around it (-) \\
        & $\cdot$ one saddle point at the resting state $(v = 0)$ \\
        & $\cdot$ one unstable focal fixed point (+) \\ \\

 $-0.2 < f < -0.0021$ & $\cdot$ one unstable focal fixed point \\
    &\, with a limit cycle around it (-) \\
        & $\cdot$ one saddle point (-)\\
         & $\cdot$ one unstable focal fixed point (+) \\ \\

 $-0.0021 \leq f \leq -0.0019$ & $\cdot$ one unstable focal fixed point \\
    &\, with limit a cycle around it (-) \\
         & $\cdot$ one saddle point (-) \\
        & $\cdot$ one unstable focal fixed point  \\
    &\, with a limit cycle around it (+) \\ \\

 $-0.0019 < f \leq 5.38$ & $\cdot$ one unstable focal fixed point \\
         & $\cdot$ one saddle point (-) \\
        & $\cdot$ one unstable focal fixed point \\
    &\, with a limit cycle around it (+) \\ \\

 $5.38 < f < 6.496$ & $\cdot$ one unstable focal fixed point (-)\\
         & $\cdot$ one saddle point (-)\\
         & $\cdot$ one stable focal fixed point (+) \\ \\

 $f \geq 6.496$ & $\cdot$ one unstable fixed point (-)\\
         & $\cdot$ one saddle point (-)\\
         & $\cdot$ one stable focal fixed point (+) \\
\hline
\end{tabular}
\end{table}

\begin{figure}
\begin{center}
\includegraphics*[width=1.0\columnwidth]{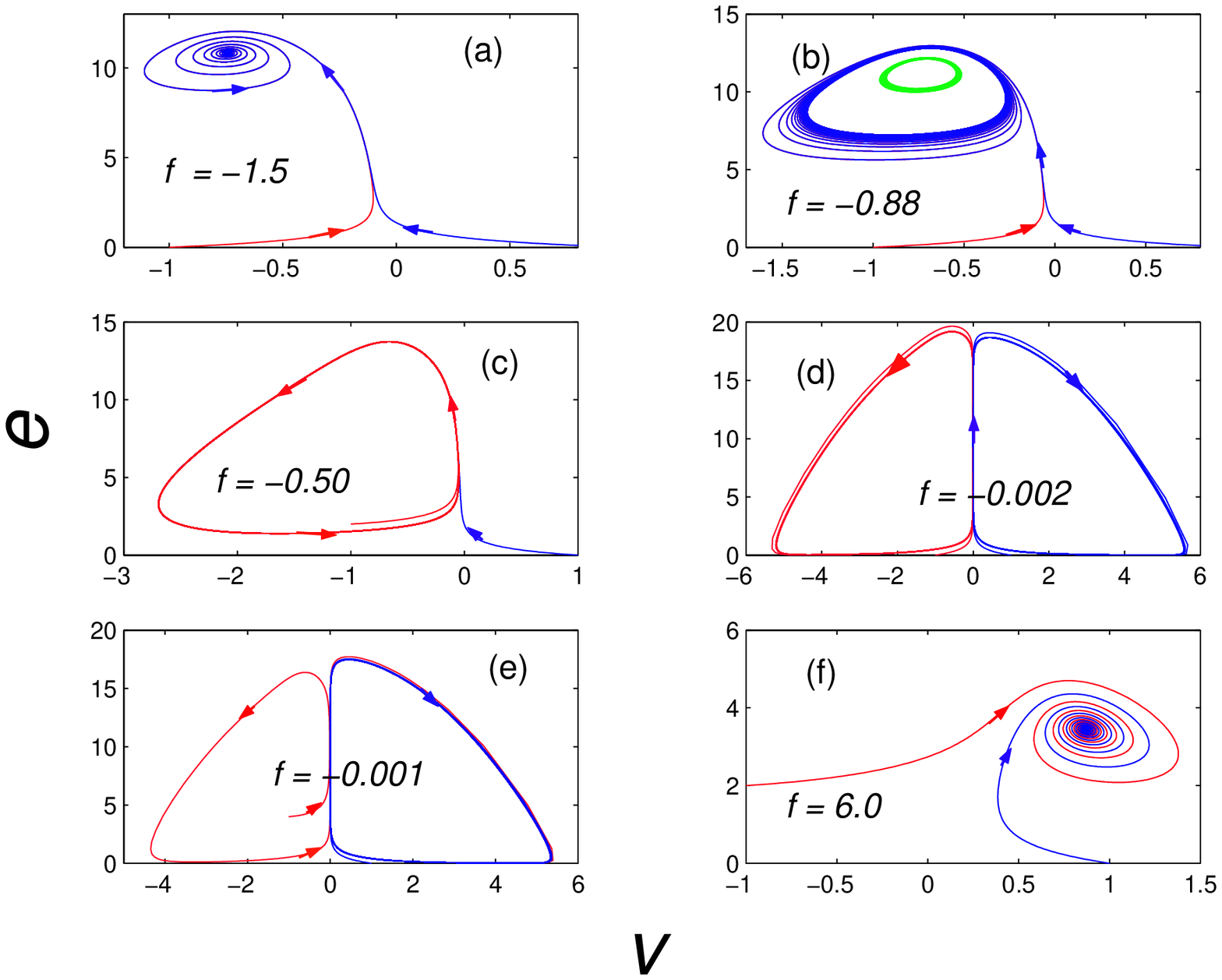}
\end{center}
\caption{(color online) Trajectories in the phase space, $\{v,e\}$
for several values of $f$ chosen from Table~\ref{T4}, when $a_{0,1}
=0.0002$, $a_{0,2} = 2.0$, $a_{0,3} = 1.2$, and $a_{0,4} = 1.0$.
(a) The particle is pulled back in a smooth way due to a heavy
load.(b) The pulled back particle moves stepwise except when it is
initialized close to the left fixed point. (c) The particle is
pulled back due to the load and moves stepwise regardless of the
initial state. (d) The particle is able to move stepwise in both
negative and positive directions depending on the initial state.
(e) The particle can overcome the load and move stepwise to the
positive direction. (f) A sufficiently large force to the positive
direction drags the particle smoothly to the positive direction
regardless of the initial state of the particle.}\label{four}
\end{figure}

When the load is large enough, $f < -0.89$, a limit cycle does not
appear, yielding a smooth movement pulled back. If
$-0.89 \leq f < -0.85$, the pulled back particle moves stepwise
except when it is initialized close to the left fixed point. On
the other hand, the particle is pulled back due to the load and
moves stepwise regardless of the initial state if $-0.85 \leq f <
-0.0021$. Note that the the saddle point moves from $v>0$ (when $f
< -0.2$) to $v<0$ (when $-0.2<f$) by crossing the value, $v=0$ when $f=-0.2$.
In a very narrow range, $-0.0021 \leq f < -0.0019$,
the particle is able to move stepwise in both negative and
positive directions depending on the initial state. When $-0.0019
< f \leq 5.38$, the particle can overcome the load and move
stepwise to the opposite direction of the load.
For a sufficiently large force
to the positive direction, the particle is dragged smoothly to the
same direction of the force regardless of the initial state of the
particle. We plot these trajectories in Fig.~\ref{four}.

\section{Stepping Motion}

The limit cycle in our phase space, $\{v,e\}$ indicates a
stepwise motion in time. Moreover a particle carrying a load
($f$, in this model) can perform a directed step-by-step motion
overcoming the load. This motion is similar to that of a
processive molecular motor in a cell. Here, we analyze an engine
mechanism of the limit cycle and relate it to the walking
mechanism of the molecular motors. A noise effect to the stepping
motion in our generalized energy depot model is also discussed.

\begin{figure}
\begin{center}
\includegraphics*[width=1.0\columnwidth]{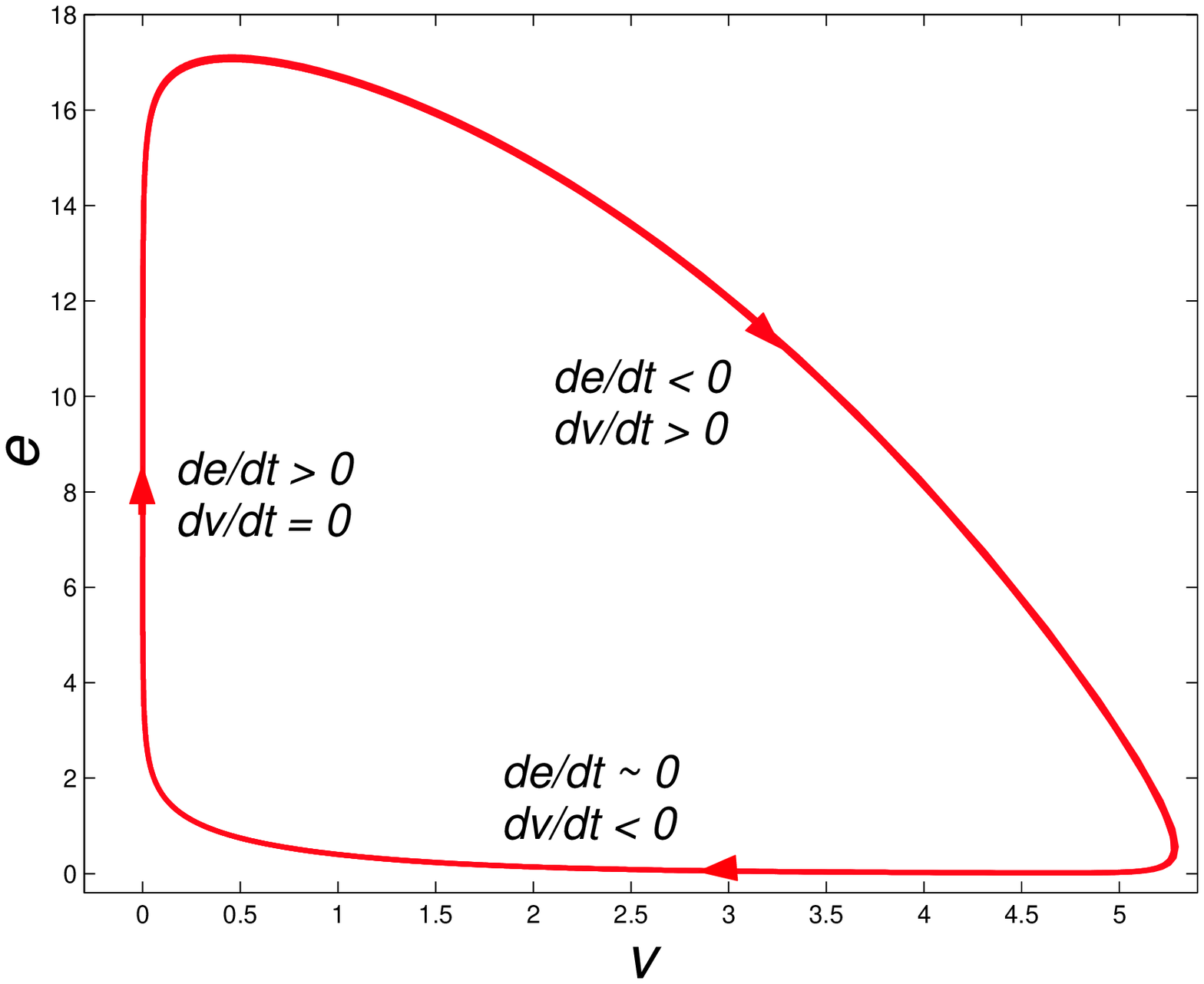}
\end{center}
\caption{The engine mechanism of the limit cycle. The motion can
be approximately divided into three states; resting ($\dot{e}> 0$,
$\dot{v} = 0$), accelerating ($\dot{e} < 0$, $\dot{v} >0$), and
decelerating($\dot{e} \sim 0$, $\dot{v} < 0$).} \label{three}
\end{figure}

Fig.~\ref{three} shows  a typical pattern of the limit cycle
appearing in our model. This triangle-like limit cycle describes
a cycling motions between three states
which are resting ($\dot{e}> 0$, $\dot{v} = 0$),
accelerating ($\dot{e} < 0$, $\dot{v} >0$), and
decelerating ($\dot{e} \sim 0$, $\dot{v} < 0$). The energy is
supplied during the resting state, consumed for the acceleration,
and barely changed during the deceleration. Hence, cyclic repetition of this
motion leads to the stepping motion as mentioned in the previous
sections.

To explain the walking mechanisms of molecular motors such as
kinesin and myosin V, a three-state model has been
proposed~\cite{PNAS_v97_9482,NATURE_v405_804}. The three-state
model describes ``ATP-binding'' to the motor at rest,
``ATP-hydrolysis and movement'', and ``attachment and ADP-releasing''.
The ``ATP-binding'' plays a role of the energy input ($\dot{e}>0$)
while stalling ($\dot{v} =0$) and the ``ATP-hydrolysis and
movement'' corresponds to the energy consumption ($\dot{e}<0$) for
the acceleration ($\dot{v}>0$) in our model. When ADP is released,
there is no energy cost. Hence the ``attachment and ADP-releasing''
state corresponds to decelerating state ($\dot{e} = 0,\dot{v} <
0$) in our model. The difference between the three state model for
the molecular motor and our generalized energy depot model is that
the molecular motors are walking on tracks such as microtubule or
actin filaments. However, the interaction between the motor protein
and the microtubule or the actin filament can be considered in our
model by introducing the corresponding external force. We will
report on this work later elsewhere.

\begin{figure}
\begin{center}
\includegraphics*[width=1.0\columnwidth]{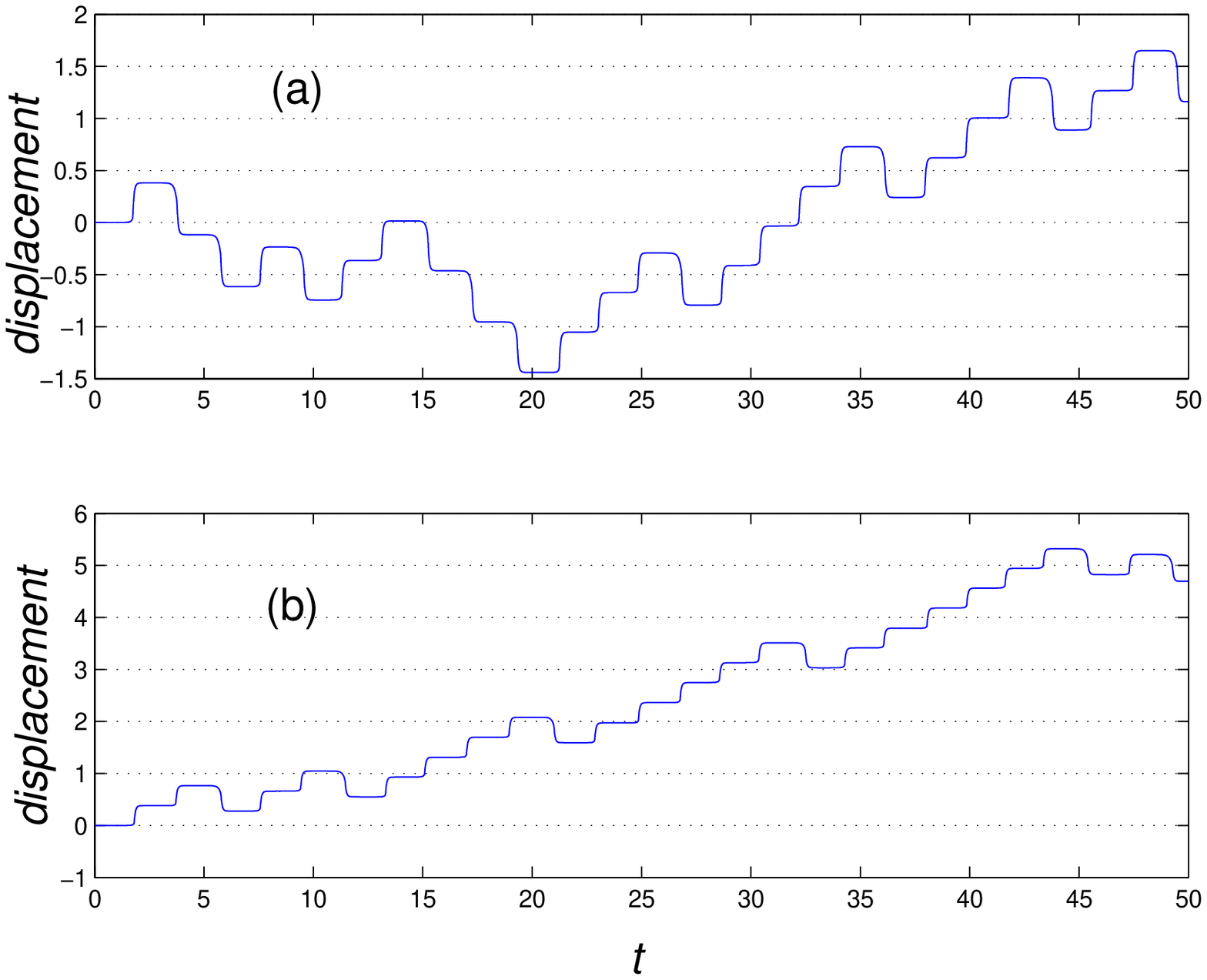}
\end{center}
\caption{(color online) The stochastic stepping motion of the
active particle with a constant load $f$ when $a_{0,1} = 0.0002$,
$a_{0,2} = 2.0$, $a_{0,3} = 1.2$, $a_{0,4} = 0.93$, and
the noise strength is $0.6$; (a) $f = -0.002$ (b) $f =
-0.001$.} \label{noise}
\end{figure}

So far, we have neglected the stochastic noise contribution to the
motion of the active particles.
However, the thermal fluctuation is inevitable and expected to
play an important role especially in the motion of the molecular
motors and nano-robots. Hence we
include the stochastic term with the noise strength $\sqrt{2k_BT\mu_0}$
in calculation of the motion.
We calculate the effect of the stochastic noise to the
stepping motions described in Fig.~\ref{four}(d) and (e) in which
the applied loads are $f=-0.002$ and $f=-0.001$, respectively. The
stochastic stepping motions are plotted in Fig.~\ref{noise}, when
the strength of the noise is $0.6$.
As expected, the particle can
move forward or backward as a consequence of the noise
fluctuation, yielding irregularly directed stepwise motion as
observed in the motion of the molecular
motors~\cite{NATURE_v365_721,Cell_v126_335,QRBP_v40_87}.
Here, we compare the present work to the existing studies on
the directed motion of active particles~\cite{PhysicaA_v273_294,
EPJB_v14_157,ActaPP_v39_1251}. In those works,
the asymmetry of space was described by an external rachet
potential~\cite{PhysicaA_v273_294,EPJB_v14_157,ActaPP_v39_1251}. However,
in our model, this asymmetry is given by the conversion function, i.e., the
energy depot itself. In our model, the noise introduces occasional backward
motion as observed in
experiments~\cite{NATURE_v365_721,Cell_v126_335,QRBP_v40_87}, unlike
previous models. Therefore, further study on the effect of the external ratchet
potential on the present model appears necessary to understand the physical
origin of the motional polarity.

\section{Negative Stiffness}

The conversion rate of the internal energy into the motion is in
general a function of space and velocity
~\cite{Schweitzer,PLB_v659_447}. As we have observed that the
velocity dependent conversion rate can induce an active behavior
in certain velocity range, it is expected that a position
dependent conversion rate may also introduce an active behavior in
oscillation. In this section, we consider a special form of the
space-dependent conversion rate to show that our generalized
energy depot model induces an active behavior in certain spatial
range. This active behavior in an oscillatory system appears as
the negative stiffness of membrane or bundle. As an example, it is
known that the spontaneous oscillation of the mechanosensitive
hair bundle in the inner ear is strongly related to the negative
stiffness of the bundle~\cite{PNAS_v102_16996}.
Therefore, we believe that it is desirable to investigate
contributions from a conversion rate function, which depends on velocity and
position simultaneously.

\begin{figure}
\begin{center}
\includegraphics*[width=1.0\columnwidth]{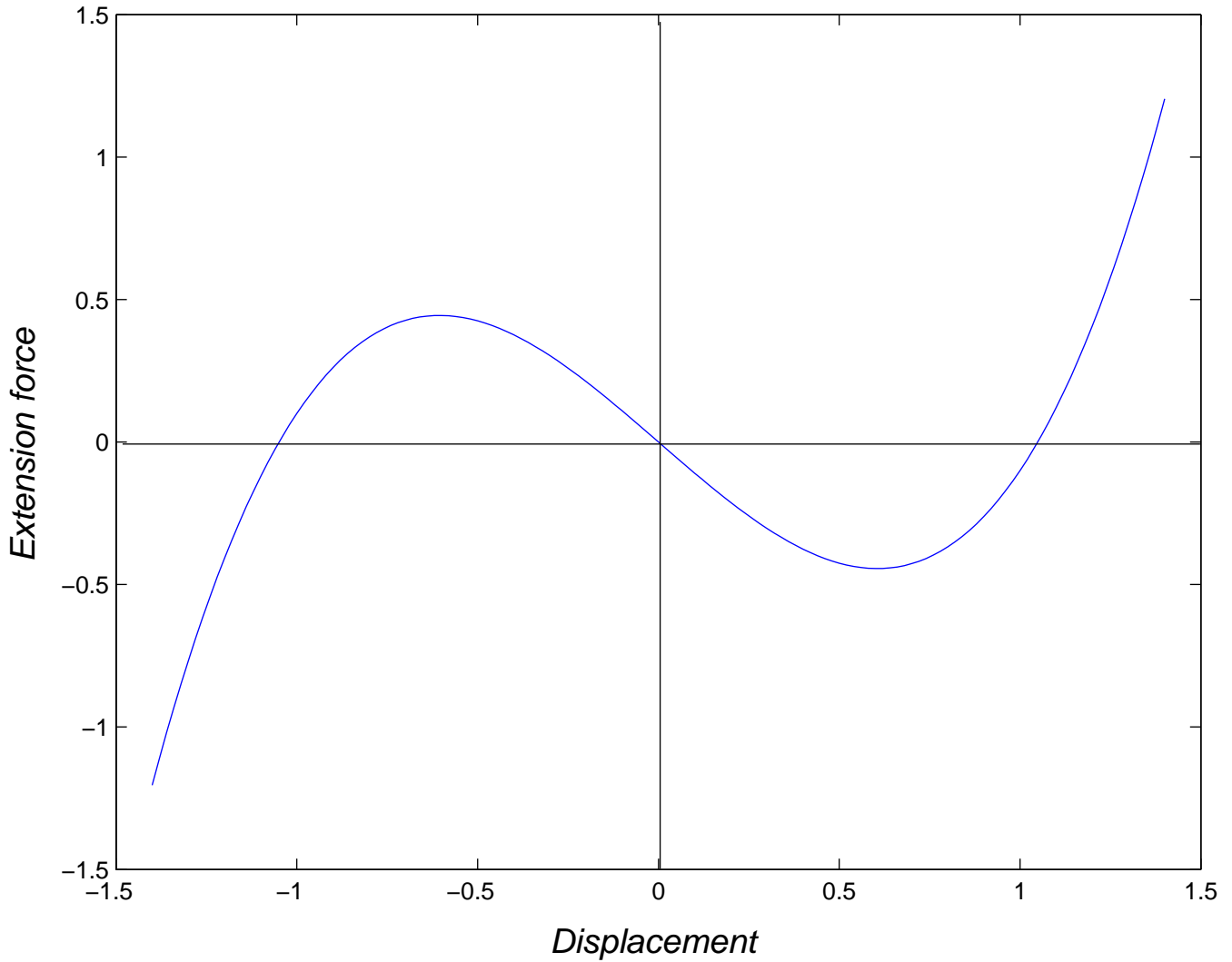}
\end{center}
\caption{Extension force $\kappa_{\mathrm{eff}}x$ versus $x$ when
$d(x,v)= a_{1,1}xv + a_{3,1}x^3v$. In this plot, we use that
$a_{1,1}=2.1$, $a_{3,1}=-1$, $\kappa = 1$, and $e=1$.} \label{negative}
\end{figure}

Motivated by this observation, we consider the following
form of the conversion rate,
\begin{eqnarray}
d(x,v) = a_{1,1}xv + a_{3,1}x^3v. \label{gf1}
\end{eqnarray}
For the stationary state, $\dot{e}=0$, the equation of motion
becomes
\begin{eqnarray}
m\dot{v} = -\mu_0 v - \kappa_{\mathrm{eff}} x + f, \label{stiff_1}
\end{eqnarray}
where the effective stiffness is
\begin{eqnarray}
\kappa_{\mathrm{eff}} = \kappa - (a_{1,1} + a_{3,1}x^2)e.
\label{stiff_2}
\end{eqnarray}
To avoid infinite amplitude of vibration, $a_{3,1}$ should be negative.
Note that if the second term exceeds the physical stiffness $\kappa$,
$\kappa_{\mathrm{eff}}$ could be negative in the region
$-\sqrt{(a_{1,1}e-\kappa)/|a_{3,1}|e} < x < \sqrt{(a_{1,1}e -
\kappa)/|a_{3,1}|e}$. Fig.~\ref{negative}
plots the extention force $\kappa_{\mathrm{eff}}x$ versus $x$.
This type of the negative stiffness has been experimentally
observed and explained using a two-state
model~\cite{PNAS_v102_16996}.  Thus, it will be interesting to investigate
possible connections between the position dependent conversion rate and
the two-energy-state model. Another possible application of the position
dependent conversion rate is introduction of a protective mechanism for
the system from excessive displacement. We believe that a suitable choice
of the conversion rate can provide such a protective mechanism for
oscillatory nano-robots.

\section{Conclusion}

In this paper, we have presented a general model of active
Brownian particle in which the conversion rate of the internal
energy into the motion is a general function of
the velocity of the particle. We have shown that when the
conversion rate depends only on $v^2$ and $v^4$, an
active amplification and a protective braking mechanism appear. We
believe that this active amplification with a braking mechanism
can provide a new paradigm for any underdamped biological systems
and nano-machines. It is conceivable that the evolutionary tactic
in biological system may have provided such protective mechanisms
to any underdamped biological units to protect them from excessive
oscillations. Such a self-adapting mechanism will be also needed
in designing nano-machines.

When the conversion rate is asymmetric in velocity, various
interesting behaviors appear. One of the most remarkable behaviors
is a directed motion without any external forces. Such a directed
motion suggest a possibility of overcoming any extra load on the
particle. Another most remarkable behavior is a stepping motion
described as a limit cycle. Such a stepping motion has a
polarity hence being applicable to the molecular motors.

When the conversion rate depends on not only the velocity but also
the position of the particle, it has been shown, for a very
special case, that the effective stiffness of an oscillatory
system could be negative. Such a behavior has been observed in the
hair bundle motion in the inner ear. Hence it also suggests a
possible connection between our generalized energy-depot model and
a relevant biological system.

We have shown that various interesting active
motions which exist in real world can be induced for active
Brownian particles by considering various forms of
the energy conversion into motion. It is quite remarkable that our
general energy-depot model could describe such a variety of
motions even though the generic relations between our model and
the real world are to be investigated further.

\section{Acknowledgment}

The authors thank to Professors Yongah Park and Myung-Hoon Chung
for fruitful discussions. This work is supported by the Korea
Science and Engineering Foundation (KOSEF) grant funded by the
Ministry of Education, Science and Engineering (MEST),
Korea(R01-2006-000-10083-0).


\end{document}